\input{aipcheck}

\documentclass[
    ,final            % use final for the camera ready runs
  ]
  {aipproc}

\layoutstyle{8x11single}

\begin{document}

\title{Gleason, Kochen-Specker, and a competition that never was}

\author{Ingemar Bengtsson}{
  address={Fysikum, Stockholms Universitet, 106 91 Stockholm, Sweden}
}


\begin{abstract}
I review the two theorems referred to in the title, and then 
suggest that it would be interesting to know how much of Hilbert 
space one can use without forcing the proof of these theorems. 
It would also be interesting to know what parts of Hilbert space 
that are essential for the proofs. I go on to discuss cubes, 
graphs, and pentagrams.
\end{abstract}

\maketitle

%%%%%%%%%%%%%%%%%%%%%%%%%%%%%%%%%%%%%%%%%%%%
%% MAINMATTER
%%%%%%%%%%%%%%%%%%%%%%%%%%%%%%%%%%%%%%%%%%%%

\section{Gleason}

Gleason's theorem \cite{Gleason} is one of the key structural theorems of physics. 
Gleason assumes that the projective Hilbert space description of quantum states 
is in place, and then proves that there exists exactly one way to introduce 
probabilities in this description, namely the usual way through a density 
matrix. More precisely each orthonormal basis is associated 
to a probability distribution which then governs projective measurements (so 
that we will tend to use the expressions ``orthonormal basis'' and ``projective 
measurement'' a little interchangeably). Here 
is one possible formulation of the theorem:

\smallskip
\noindent {\bf Gleason's theorem}. {\sl In a real or complex Hilbert space 
of dimension $N > 2$, let each ray $|\psi\rangle \langle \psi |$ be assigned 
a real number $p_\psi \geq 0$, such that}

\begin{equation} \sum_{i = 1}^N p_{e_i}(|e_i\rangle \langle e_i|) = 1 
\label{ett} \end{equation}

\noindent {\sl whenever the vectors $\{ |e_i\rangle\}_{i = 1}^N$ form an 
orthonormal basis. Then there exists a density matrix $\rho$ such that} 

\begin{equation} p_\psi = \mbox{Tr}\rho |\psi \rangle \langle \psi | 
\end{equation}   

\noindent {\sl for all unit vectors $|\psi\rangle $.}
\smallskip

\noindent A ray is a one-dimensional subspace, spanned by a single vector 
and represented by a projector.  
The real valued function $p_\psi$ is called a frame function. 
There are no further assumptions, in particular continuity is proved, not 
assumed. For this reason the proof of Gleason's theorem is famously difficult. 
At the same time it should be stressed that from the point of view 
of physics the assumptions are quite strong. It is assumed that the number 
$p_\psi$ remains the same regardless of which, out of many, orthonormal 
bases it is regarded as being part of. The technical term for this is that 
the probabilities are ``non-contextual''. Perhaps---as stressed by Bell 
\cite{Bell, Mermin}---this need not be the case? In any case the theorem 
highlights how much one can get out of this assumption.

The theorem dates from 1957. It is worth noticing that had it been a little 
weaker, so that we were left with a one parameter family of ways in which 
probability distributions could be extracted from the formalism, then we 
would have seen a very large number of experiments trying to set bounds on 
that parameter. The theorem has figured prominently in the literature on the 
foundations of quantum mechanics \cite{Piron}, and it is possible to dig 
deeper \cite{Pekka}, but my purpose here is a little orthogonal to what has gone 
on before.

\section{Kochen-Specker} 

Gleason's theorem has a corollary with an independent proof \cite{ES, KS}:

\smallskip
\noindent {\bf The Kochen--Specker theorem}. {\sl It is impossible to assign 
all rays of an $N > 2$ dimensional Hilbert space a truth value (true or false), 
in such a way that exactly one vector in each orthonormal basis is true.}
\smallskip

\noindent The second theorem is a corollary of the first, because if we 
replace true and false by one and zero, we have an example of a frame 
function that obeys Gleason's conditions. But probability 
distributions that come from density matrices are not of this form. It is 
a very interesting corollary, since the outcome of a projective measurement 
does provide such truth value assignments to the vectors in the given 
orthonormal basis. Apparently then these assignments did not exist prior to 
the decision to perform that particular measurement. To the extent that one can 
talk about them all, they are necessarily 
contextual even though the probabilities are not.  

Ernst Specker's original motivation concerned a problem in theology: can 
God know the outcome of all events, also those that could have happened but 
in fact did not? The answer from quantum mechanics is a clear no.  
This is usually stated somewhat less cosmically as a colouring problem. 
By convention red corresponds to true, and green corresponds to false.  
The theorem then says that it is impossible to colour the rays of Hilbert space 
using two colours only consistently with the rule derived from eq. (\ref{ett}), 
namely that exactly one vector in each orthonormal basis must be coloured 
red. 

The proof is famously easy, and proceeds by exhibiting a finite number 
of uncolourable vectors, including some orthonormal bases. In their original 
proof Kochen and Specker used a set of 117 rays for this purpose. Naturally 
this led to a competition: who can prove the Kochen-Specker theorem with the 
smallest number of rays? The upshot in dimension 3 was that Sch\"utte \cite{Peres}, 
Peres \cite{Peres}, and Penrose \cite{Penrose} produced uncolourable sets with 
33 rays, while Conway and Kochen produced one with 31 rays \cite{Peres}. In four 
dimensions Peres produced an elegant set with 24 \cite{Peres24}, Kernaghan one with 20 
\cite{Kernaghan}, and Cabello et al \cite{CEG}
a set with 18 rays. Exhaustive computer searches \cite{Pavicic} have since confirmed that 
Cabello et al are the winners. 

Several of these finite sets form seducingly beautiful configurations. While 
this is incidental to our understanding of quantum mechanics, it does add 
to the charm of the subject. We will go into the details of the Peres and 
Penrose sets later. It is perhaps worth mentioning that Peres' set of 24 rays 
form two triplets of mutually unbiased bases. Other sets involving 
such bases have been produced subsequently, systematically in four \cite{Waegell} 
and also in eight \cite{Mermin, Waegell2} dimensions. The vectors involved in 
these constructions are eigenvectors of operators belonging to a Weyl-Heisenberg 
group, and in dimensions not equal to a power of a prime number things work 
out differently \cite{Planat}---although in dimension three the 
Weyl-Heisenberg group does have something special to offer \cite{BBC}.  

\section{The competition that never was}

Given that it is impossible to colour all of Hilbert space according to the 
Kochen-Specker rules we ask: how much of it can be coloured? As rules, we choose 
to require that no two orthogonal vectors are red, and no orthonormal 
basis has all its vectors green, but it is allowed to leave some vectors 
uncoloured. Moreover it is understood 
that the red and green regions should be measurable, and the uncoloured region 
(that must perforce exist) should be measurable too.  

As far as I know only one paper, by Appleby \cite{Appleby}, has been devoted 
to this question. It confined itself to three dimensional real Hilbert space, 
so colouring the unit vectors is equivalent to colouring a 2-sphere (or 
more precisely a real projective 2-space). One of its purposes was to provide a 
lower bound for the area that must be left uncoloured. By wiggling one of the 
uncolourable configurations employed to prove the Kochen-Specker theorem a 
lower bound of around one percent was achieved. It was also noted that 87 
percent of the sphere can be coloured consistently with the rules. The idea is 
to start with a vector and colour it red, and then colour all vectors orthogonal 
to it green. Pictorially the North Pole is red and the equator is green. Then 
one enlarges the North Pole to a red polar cap, keeping it small enough so 
that one cannot place two orthogonal vectors there, and at the same time 
one creates a green belt around the equator, keeping it small enough so that 
it remains impossible to place a complete orthonormal set of vectors there. 
The polar cap then extends down to a latitude of 45 degrees, and the belt 
goes up to a latitude of 35 degrees. The sum of the area of the cap and the 
belt turns out to be 87 \% of the total area of the sphere. 

This idea can be extended to real Hilbert spaces of arbitrary dimension 
\cite{Granstrom}. Interestingly the end result is not at all obvious, since 
it depends on a delicate balance between 
two opposing effects: the width of the green belt shrinks as the dimension grows, but at 
the same time the volume of the sphere becomes concentrated to a region 
around the equator. As a result the fraction of the sphere that can be 
coloured in this way shrinks down to 67 \% (for dimension 12), and then 
raises slowly to its limiting value 

\begin{equation} \mu_\infty = \mbox{erf} \left( \frac{1}{\sqrt{2}}\right) 
\approx 0.68 \ . \end{equation}

\noindent A similar strategy can be followed for complex Hilbert spaces of 
arbitrary dimension $N$. We write a unit vector modulo phases as 

\begin{equation} \psi = (\sqrt{p_0}, \sqrt{p_1}e^{i\nu_1}, \dots , 
\sqrt{p_{N-1}}e^{i\nu_{N-1}})^{\rm T} \ , \hspace{8mm} \sum_{i=0}^{N-1}p_i = 1 \ . \end{equation}

\noindent These coordinates are very convenient for volume calculations using 
the Fubini-Study measure, since they make projective Hilbert space behave like a 
product of a flat simplex and a flat torus of a standard size. We can now 
make a red polar cap including 

\begin{equation} 1 \geq p_0 > \frac{1}{2} \ . \end{equation}

\noindent We can also make a green belt extending between  

\begin{equation} 0 \leq p_0 < \frac{1}{N} \ . \end{equation}

\noindent Using the Fubini-Study measure to compute volumes one finds that 
this colouring covers a fraction of projective Hilbert space which is 

\begin{equation} \mu_N = 1 - \left( 1 - \frac{1}{N}\right)^{N-1} + 
\left( \frac{1}{2}\right)^{N-1} \ . \end{equation}

\noindent For $N = 3$ this is 81 \%, slightly less than in the real case. It reaches a minimum of 61 \% for $N = 9$, and then raises 
slowly towards the limit 

\begin{equation} \lim \mu_\infty = 1- \lim_{N\rightarrow \infty} \left( 1 - 
\frac{1}{N}\right)^{N-1} = 1- \frac{1}{e} \approx 0.63 \ . \end{equation}

\noindent This result was never published since at this point its author 
decided she did not want to be a physicist, she wanted to be a 
poet. To those who can read Swedish I can recommend {\it Os\"akerhetsrelationen}, 
a book-length poem on the foundations of quantum mechanics \cite{Helena}. 

But as a result no attempt to beat Appleby's construction, and to colour 
more of Hilbert space than he did, has been made. I think people ought to 
compete with him. A list of attempts to do better than him might well provide 
interesting food for thought, in many ways even more interesting than the 
list of examples of finite sets of uncolourable vectors. Actually my 
suspicion is that Appleby's construction may well be optimal at least in high 
dimensions, but nevertheless such a competition might shed light on the next 
question.

Of course one can adopt other rules of the game if one wants to. 
In particular, Granstr\"om asked what fraction of all orthonormal bases that 
can be fully coloured according to the Kochen-Specker rules. Using Appleby's 
colouring she found that percentage to be 69 \% in the real three 
dimensional case, and 34 \% in the real four dimensional case \cite{Granstrom}. Possibly this fraction does go to zero as the dimension increases?

\section{What can you colour?}

It is striking that most of the known finite sets of uncolourable vectors, 
in particular all the ones with close to the minimal number of entries except 
the Penrose set, contain real vectors only. In four dimensions there is a 
definite meaning attached to this: it is known that given a set of real 
vectors one can find a tensor product decomposition such that all of them 
are maximally entangled, and conversely given a maximally entangled set of 
vectors it is possible to choose a magical basis such that they become real 
\cite{Wootters}. There is a similar interpretation of real vectors in three 
dimensions: they are maximally non-coherent with respect to a definite definition 
of spin coherent states \cite{Klyachko}. So we already know 
that the set of all maximally entangled states is uncolourable. 

The set of separable two-qubit states is colourable however. To see this, note 
that any separable state $|\psi_A \rangle \otimes |\psi_B\rangle$ can be 
coordinatized by  

\begin{equation} |\psi_A\rangle = \cos{\frac{\theta_A}{2}}|0\rangle + 
\sin{\frac{\theta_A}{2}}e^{i\phi_A}|1\rangle  \ , \end{equation}

\noindent and similarly for $|\psi_B\rangle$. Now divide the set of separable 
states into four sets, 

\begin{eqnarray} {\rm I} : \ (\phi_A,\phi_B) \in [0,\pi)\times [0,\pi) \ , 
\hspace{10mm} 
{\rm II}: \ (\phi_A,\phi_B) \in [0,\pi)\times [\pi,2\pi) \ , \ \ \nonumber \\ 
\ \\ 
{\rm III} : \ (\phi_A,\phi_B) \in [\pi ,2\pi)\times [0,\pi) \ , 
\hspace{6mm} 
{\rm IV}: \ (\phi_A,\phi_B) \in [\pi, 2\pi)\times [\pi,2\pi) \ . \nonumber  \end{eqnarray} 

\noindent There are no orthogonalities within these sets, so we can colour 
the first set red and the other sets blue.  

This suggests an alternative strategy for colouring a four dimensional 
Hilbert space, namely to start by colouring the separable states and then 
to work outwards from there. Will this beat Appleby's strategy? Will it 
lead to a larger fraction of coloured bases?

\section{A comment on impossible cubes}

The Penrose set of 33 vectors is special in that it cannot be made real by 
any choice of basis. This goes a little against the drift of the idea that 
it should be the real vectors that resist colouring the most, but on closer 
inspection one finds that this set is complex only in a mild way. Let us recall 
how the Penrose set, and that of Peres, arises. In his print {\it The Waterfall}, M. C. 
Esher illustrates one of the Penrose's impossible pictures \cite{Lionel}. 
Of concern for the moment are the three interlocking cubes that grace the top 
of one of the towers in Escher's print. We observe that a cube is naturally 
associated to 3 + 6 + 4 = 13 rays, 
three going through the middle of a face, six through the middle of an edge, 
and four connecting the corners. This gives rise to an interesting orthogonality 
graph with 13 vertices representing the rays. In such a graph each vertex represents 
a vector, and the vertices are connected with a link if and only if they represent 
orthogonal vectors. (The converse problem, to find the lowest dimension in which a 
given graph can be realized as an orthogonality graph, is non-trivial.) The Kochen-Specker 
rules ask us to colour the vertices of such a graph red or green, subject to 
the rules that two linked vertices cannot both be red, and any complete 
triangle must contain exactly one red vertex (if we are in three dimensions, 
where three mutually orthogonal vectors form a basis). The cube graph is colourable---but 
still very interesting in that it leads to a state-independent probabilistic proof 
of the Kochen-Specker theorem. This was discovered recently by Yu and Oh \cite{YO}, 
and discussed in Larsson's talk, but I will not go into this now. I just observe 
that in this connection the graph is a 
way to visualize the cube, and I also observe that the vectors are determined 
uniquely up to an overall unitary transformation by the orthogonalities encoded 
in the graph.  

The impossible cubes that appear in Escher's print arise if you start with one 
cube, and add two cubes obtained by rotating the first cube ninety degrees around 
an axis that connects two opposing edges. A pair of two orthogonal axes of this 
kind is needed to produce the three interlocking cubes. Each new cube gives rise 
to ten new vectors, while three vectors are shared by all three cubes, so the total 
number of vectors appearing in the print equals 33. Six intercube orthogonalities 
arise, and this turns out to be just enough to make the completed orthogonality 
graph uncolourable. 

\begin{figure}[ht]
  \resizebox{15pc}{!}{\includegraphics{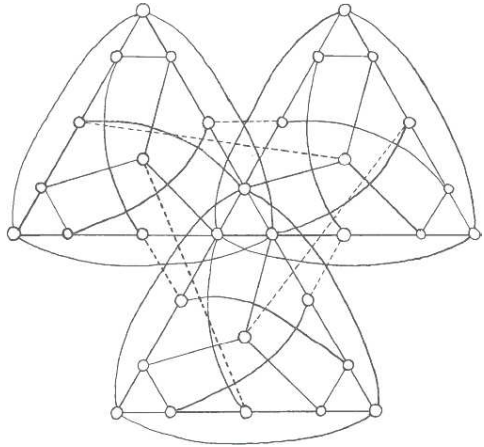}}
\caption{The Escher-Peres interlocking cubes. Each 13-vertex triangle is a copy 
of the Yu and Oh graph, including 24 orthogonalities given by the links. Reproduced 
with permission \cite{KB}.}
\end{figure}

In the basis provided by the three vectors common to all three cubes the second 
cube is obtained from the first by 

\begin{equation} R_{I\rightarrow II} = \left( \begin{array}{ccc} e^{i\phi} 
& 0 & 0 \\ 0 & 0 & -1 \\ 0 & 1 & 0 \end{array} \right) \end{equation} 

\noindent and the third from the first by  

\begin{equation} R_{I\rightarrow III} = \left( \begin{array}{ccc} 0  
& 0 & -1 \\ 0 & e^{-i\phi} & 0 \\ 1 & 0 & 0 \end{array} \right) \end{equation} 

\noindent ---and we took the liberty to add an undetermined phase factor, since 
this leaves all the orthogonality relations in place. The complex Penrose set can 
now be obtained from the Peres set by a suitable choice of this free phase 
\cite{Gould}. It seems right to say that the 
former is only mildly complex. Incidentally the fact that there is a free phase 
in the realization of the orthogonality graph seems to be unusual, given that 
the graph is sufficiently involved to give a proof of the Kochen-Specker theorem 
\cite{Kate}.   
 
The remainder of Escher's print illustrates a different holistic mathematical 
concept and is of no concern here, but it is clearly very appropriate for his 
print to include a proof of the Kochen-Specker theorem. 

\section{Motivation}

After this pleasant excursion into the world of Art, let us return to Science. 
Does this circle of ideas have experimental consequences? In fact it has. To 
see this let us focus on a simple subgraph of the orthogonality graphs that 
we have encountered, namely the pentagon (or equivalently the pentagram). This 
is in fact a very interesting graph in itself, as was noted by Wright \cite{Wright} 
and Klyachko et al. \cite{Can}. Moreover it occurs as a subgraph in the description 
of several ``paradoxical'' quantum arguments that have attracted the attention 
of experimentalists, including Hardy's paradox \cite{Piotr} and the CHSH 
inequality \cite{Severini, Sadiq}. It 
is certainly colourable, but it should be observed that at most two vertices 
can be coloured red, meaning that at most two of the potential outcomes can 
come out as ``true''. 

Now consider a classical experiment, with meaurements $P_i$ whose outcomes 
obey the Kochen-Specker rules. We can imagine a platter with a 
pentagram drawn on it, and five cups placed upside down over the vertices. 
Each time the platter is presented to us we look under two cups connected by 
a link, and record the number of stones (1 or 0) found under each cup. It 
never happens that there are stones under both cups, and from this 
fact---assuming that our 
choice of which pair of cups we look under has nothing to do with the 
preparation of the platter---we reason that the total number of stones 
on the platter never exceeds 2. After many trials we then expect  

\begin{equation} \sum_{i=1}^5\langle P_i\rangle \leq 2 \ . \label{Klyachko} 
\end{equation}

\noindent But in a corresponding quantum experiment the measurements 
correspond to projectors $\hat{P}_i$ with orthogonality properties given 
by the graph. The latter restrict the spectrum of the operator 

\begin{equation} \Sigma = \sum_{i=1}^5\hat{P}_i \end{equation}

\noindent in such a way that the maximal eigenvalue of $\Sigma$ never exceeds 
$\sqrt{5}$ \cite{Klyachko, Piotr}. We therefore find the quantum bound

\begin{equation} \sum_{i=1}^5\langle \psi|\hat{P}_i|\psi \rangle = 
\langle \psi |\Sigma |\psi \rangle \leq \sqrt{5} \ . \end{equation}

\noindent We can exceed this bound with the classical platter if it 
is prepared in 
a conspiratorial way: only one stone is used in the preparation, but it 
is always placed under one of the two cups we will decide to look under. 
Thus we get 

\begin{equation} \mbox{classical bound} = 2 < \mbox{quantum bound} = \sqrt{5} 
< \mbox{conspiratorial bound} = 2.5 \ . \end{equation}

\noindent The Yu and Oh set of vectors (associated to the cube described earlier) lead 
to a similar inequality, but---like some other sets introduced earlier \cite{AC}---with 
the interesting difference that the projectors form an equal weights POVM, so 
that the classical bound is violated equally by all quantum states. 

These bounds are testable. In fact they have been tested, and 
we will come back to this. In principle quantum mechanics is doubly on the line 
in such an experiment: the classical inequality could hold, or the quantum bound 
could be exceeded. 
 
For systems whose Hilbert space dimension does not exceed four so many 
transformations have been done in the lab that it is completely unreasonable 
to expect any problems with the quantum mechanical description. However, from 
some points of view higher dimensional Hilbert spaces are very mysterious. In 
particular they are that from a point of view that I myself often adopt, that 
of Mutually Unbiased Bases (MUBs) and Symmetric Informationally Complete POVMs (SICs). 
There does exist a research program aiming to reconstruct Hilbert space from 
SICs \cite{Fuchs}, but it is conceivable that SICs do not exist in dimension 77 
(say). Does this mean that such research programs run the risk of reconstructing 
something that is only approximately a Hilbert space, and if so what do we do? 
Is it obvious that we will stick to Hilbert space? For MUBs these 
questions arise already in dimension 6. 

Some of the pioneers in testing the Bell inequalities did expect them to hold, 
and quantum mechanics to fail \cite{Freire}. Nobody expects this to happen in 
a pentagram experiment, but I am suggesting that we have as yet no reason to be 
dogmatic about the outcome of similar experiments for higher dimensional systems.  
 
\section{Remarks on graphs}

So far I have not even hinted whether the construction of ``paradoxical'' graphs is 
an art or a science. Actually it turned into a science quite recently, and it is 
by now known what properties a graph must have in order to give rise to 
Kochen-Specker inequalities of the kind I have discussed. In particular, 
what we have called the classical, the quantum, and the conspiratorial bounds 
turn out to be identical with quantities that have been studied by graph 
theorists  \cite{Severini}. They are known respectively as the independence 
number (defined as the 
maximum number of pairwise disconnected vertices in the graph), the Lov\'asz 
theta-function \cite{Lovasz}, and the fractional packing number. The Lov\'asz 
theta-function can be obtained by semi-definite programming, so the quantum bound 
can be effectively computed. This is interesting because the problem of finding 
the classical bound---while easy enough for small graphs---turns out to be NP complete. 
This cross-disciplinary discovery have 
made some powerful tools available to the quantum theorist, and many 
developments stem from here. It would take us too far afield to try to 
account for them here. 

One important point, that I should mention, is that many recent 
experiments are aiming not at the inequality (\ref{Klyachko}) (and its 
relatives for other graphs) but at correlation inequalities that can be 
derived without reference to the Kochen-Specker rules. They depend only on 
the assumption of pre-determined outcomes \cite{Can}, but again I will not go 
into this here.

\section{Experiments}

So far two quantum optical experiments have been made to test the pentagram inequality 
\cite{Lapkiewicz, Ahrens}. 
As soon as one looks at such an experiment in the lab, one begins to appreciate 
the force of the comment that ``the result of an experiment may reasonably 
depend not only on the state of the system ... but also on the complete 
disposition of the apparatus''. This quote is not from Bohr, it is Bell 
\cite{Bell} questioning the assumption that a hidden variables theory must 
be non-contextual, or whether the outcome of the measurement associated to 
$P_2$ might depend on which of the mutually incompatible measurements 
$P_1$ and $P_3$ it is measured together with (given that it is compatible 
with both). 

Both experiments prepare a state that should lead to a maximal violation 
of the classical bound, and the results are in quite good agreement with 
what one would expect. Still, though beautiful, the experiments do fall 
short of the ideal. They perform 
compatible measurements---say ten thousand measurements---in pairs, beginning 
with $P_1$ and $P_2$, going on to $P_2$ and $P_3$, and so on, ending with 
$P_5$ and $P_1$. It would be desirable to let a random generator decide each 
time what pair of measurements that is to be performed, in order to stymie 
any conspiratorial preparation of the classical platter. In the first experiments 
there is also the difficulty that the measurement performed together with 
$P_5$ is not in fact identical to the measurement $P_1$, and special 
measures must be taken to deal with this. In effect one of the vertices of 
the pentagram does not close. You may recall that it was precisely through such 
an open vertex that the devil got into Faust's pentagram \cite{Goethe}. 

\begin{figure}[ht]
  \resizebox{15pc}{!}{\includegraphics{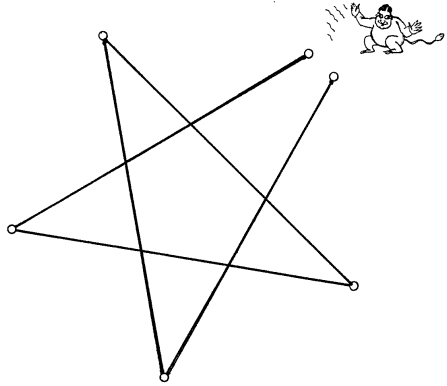}}
\caption{An experimental afterthought. With apologies 
to George Gamow, and to the organizers of the Pauli session.}
\end{figure}

There is room for further refinements of these experiments. I expect that 
many years will pass before we can listen to a talk entitled ``towards a 
loop-hole free contextuality test''. Still the refinements are important, 
and will serve as preparation for the day when such an experiment can be 
done for six dimensional systems---where quantum mechanics just possibly 
may begin to crack.
 
\begin{theacknowledgments}
I thank Kate Blanchfield for help with the cubes, and two well known colleagues 
for planting the idea that tests of 
quantum mechanics in higher dimensions may surprise us. That Andrei 
Khrennikov arranged a successful conference did not surprise me at all. 

\end{theacknowledgments}

%Mitt eget krafs

%% If the bibliography is produced without BibTeX comment out the

\bibliographystyle{aipproc}   % if natbib is available

\bibliography{sample}

\IfFileExists{\jobname.bbl}{}
 {\typeout{}
  \typeout{******************************************}
  \typeout{** Please run "bibtex \jobname" to optain}
  \typeout{** the bibliography and then re-run LaTeX}
  \typeout{** twice to fix the references!}
  \typeout{******************************************}
  \typeout{}
 }

\end{document}